\documentclass[%
 reprint,
superscriptaddress,
showpacs,
 amsmath,amssymb,
 aps
]{revtex4-1}

\usepackage{graphicx}
\usepackage{dcolumn}
\usepackage{bm}
\usepackage{color}
\usepackage{here}

\usepackage{url,listings}

\begin{document}
\title{Model-Free Cluster Analysis of Physical Property Data using Information Maximizing Self-Argument Training}
\author {Ryohto Sawada}
\affiliation{System Platform Research Laboratories, NEC Corporation, Tsukuba 305-8501, Japan}
\author{Yuma Iwasaki}
\affiliation{System Platform Research Laboratories, NEC Corporation, Tsukuba 305-8501, Japan}
\affiliation{JST, PRESTO, Saitama, 332-0012, Japan}
\author{Masahiko Ishida}
\affiliation{System Platform Research Laboratories, NEC Corporation, Tsukuba 305-8501, Japan}

\begin{abstract}
We present the semi-supervised IMSAT, a versatile classification method that works without labeled data and can be tuned by little additional information. We demonstrate how semi-supervised IMSAT can classify XRD patterns and thermoelectric hysteresis curves in the same way even though their shape and dimensions are different. Our algorithm will accelerate automation of big data collection and open a way to study artificial intelligent driven material development.
\end{abstract}

\maketitle

\section{Introduction}
Big data analysis and machine learning are being applied in many fields of fundamental sciences, with physics being no exception \cite{behler2007, matthias2012, iwasaki2017, elia2017, carleo2017}. In the field of material development, big data collection is an extreme bottleneck. Therefore, high-throughput materials fabrication and characterization are in strong demand\cite{takeuti2005, ludwig2008}. 

Composition-spread experiments are one promising solution where one can fabricate the gradient of composition in a single fabrication. For example, Yoo et al, fabricated a Fe-Ni-Co ternary alloy and measured a continuous phase diagram \cite{yoo2006} and Wang et al, fabricated La$_{1-x}$(Ca, RE)$_{x}$VO$_{3}$ and measured thermoelectricity \cite{wang2013}.
 
However, raw experimental data is usually too noisy and verbose to analyze. For example, raw experimental data varies depending on the experimental system (e.g. power of the source and sensitivity of detector) in the case of X-ray diffraction(XRD). Therefore, one needs to process the spectrum into a crystal structure for analysis. Such clustering is usually carried out by hand. Therefore, automation of clustering is in great demand to decrease costs and achieve success. 

Iwasaki et al., demonstrated clustering of the X-ray diffraction data of Fe-Co-Ni ternary-alloy thin film using normalized and constrained dynamic time warping (NC-DTW) \cite{iwasaki2017}. The key question for the automated clustering is how to quantify the similarity between two pieces of data. In the case of XRD, the spectrum is given as $s(x)$ where $x$ is the diffraction angle. The similarity between the two pieces of data $s,t$ is defined by kernel function $D(s,t)$. 
In the case of NC-DTW, $D(s,t)$ is given by
\begin{eqnarray}
D(s,t) = \min \sum_{i}^{N} (s'_{j(i)} -t'_{i})^{2}
\end{eqnarray}
where
\begin{eqnarray}
s' = \frac{s}{(\sum_{i}^{N} s_{i}^{2})^{1/2}}, t'=\frac{s}{(\sum_{i}^{N} t_{i}^{2})^{1/2}}
\end{eqnarray}
and $j(i)$ must satisfy
\begin{eqnarray}
i \leq i' \Rightarrow j(i) \leq j(i'),\  |j(i) - i | < w, \nonumber \\
 j(1) = 1,\  j(N) = N.
\end{eqnarray}
where $w$ is the window size that limits the range of time warping. Iwasaki et al. also tried different kernel functions (e.g. Euclidean, Manhattan, Pearson and cosine) and found that other kernel functions cannot classify a crystal structure because they cannot accommodate peak shifting due to lattice constant change. 

However, the appropriate kernel function varies depending on the problem. Furthermore, many of existing kernel functions, including NC-DTW, are limited to low dimensional classification although a lot of raw experimental data is complicated multi-dimensional data. These problems prevent us from reusing kernel functions and make the automation non-profitable.

A neural network is a promising approach to achieve versatility. Compared to the previous approaches, applying neural network to multidimensional data. One can solve various problems using the same neural network  e.g. image recognition, text recognition and sound recognition \cite{bishop2010, hope2017, osinga2018}, chaotic phase and quantum mechanics \cite{behler2007, elia2017, carleo2017}. However, most of the previous applications use supervised training and supervised training for pre-clustering is costly because supervised training generally requires a large amount of data.

In this paper, we present a comprehensive solution based on information maximizing self-argument training (IMSAT) \cite{hu2017} that does not require hands-on searches of kernel functions or preparation of large amount of data for supervised learning. We have demonstrated our algorithm succeeds in clustering line charts and scatter plots from raw experimental data. Our algorithm can accelerate automation of big data collection and open a way to study artificial intelligent driven material development.

\section{Methods}

Model complexity is the origin of the versatility of a neural network; however, it is also the reason why a neural network can easily overfit small data sets. Therefore, the degree of freedom of the neural network  has needs to be reduced in order to avoid overfitting by "regularization". 
Recently, neural network regularized by local perturbation succeeded in clustering handwritten numerals with only a small amount of data. 
Virtual adversarial training (VAT) \cite{miyato2015} is a representative regularization methods based on local perturbations.  
The objective function of VAT is defined by following function: 
\begin{eqnarray}
R_{vat}(\theta) &=& R_{pert}(\theta) + H_{l}(\theta)
\end{eqnarray}
where
\begin{eqnarray}
R_{pert}(\theta) &=& \sum_{i}^{N} ( -\sum_{y'}^{V_{y}} p_{\theta}(y' | x_{i}) \log _{p_{\theta}} (y' | T_{\theta}(x_{i}))) \nonumber \\
H_{l}(\theta) &=& \beta ( \sum_{j}^{N^{l}} \log p_{\theta}(y^{l}_{j} | x^{l}_{j}), \nonumber
\label{eq:vat}
\end{eqnarray}
$\theta$ is parameter of the neural network, $N$ is the number of data, $x_{i}$ is $i$-th data, $V_{y}$ is the number of clusters, $p(y|x)$ is conditional probability, $T_{\theta}(x_{i})$ is perturbated data, $N_{l}$ is the number of data with label information, and $\beta$ is hyper parameter. $H_{l}$ is the same as the target function of supervised learning. 
$T_{\theta}(x_{i})$ is chosen to be 
\begin{eqnarray}
T_{\theta}(x) &=& \arg \max _{x'} R_{vat} (\theta; x, x') \nonumber \\
&=& \arg \max_{x'} -\sum_{y'}^{V_{y}} p_{\theta}(y' | x_{i}) \log _{p_{\theta}} (y' | x').
\end{eqnarray}
The regularization using local perturbation is based on the idea that it is preferable for data representations to be locally invariant (i.e., remain unchanged under local perturbations on data points). The idea would help neural network to learn meaningful representation of data.

Information maximizing self-argument training (IMSAT) is an expansion of VAT for unsupervised learning. Objective function of IMSAT is defined by the following equation: 
\begin{eqnarray}
R_{pert}(\theta) - \lambda (\mu H(y) - H(y|x))
\end{eqnarray}
where $\mu$ and $\lambda$ are hyper parameters, $H(y)$ and $H(y|x)$ are marginal entropy and conditional entropy
\begin{eqnarray}
H(y) = h(\frac{1}{N} (\sum_{i}^{N} p_{\theta}(y|x_{i}))
\end{eqnarray}
\begin{eqnarray}
H(y|x) = \frac{1}{N} \sum_{i}^{N} h(p_{\theta}(y|x_{i}))
\end{eqnarray}
and $h(p_{\theta}(y | x))$ is the entropy function
\begin{eqnarray}
h(p_{\theta}(y)) = -\sum_{y'} p_{\theta}(y') \log ( p_{\theta}(y') ).
\end{eqnarray}
Increasing the marginal entropy $H(y)$ encourages uniformity among the cluster sizes, while decreasing the conditional entropy $H(y|x)$ encourages unambiguous cluster assignments. IMSAT achieved over 90\% accuracy in unsupervised learning of tje clustering of handwritten numerals.

The original IMSAT is not suitable for regarding specific differences as important because IMSAT just tries to make data representation locally invariant. However, specific difference are sometimes regarded as important due to domain knowledge. Therefore, we added $H_{I}$ to enable semi-supervised learning. Our algorithm optimizes the following function: 
\begin{eqnarray}
R_{vat}(\theta) - \lambda (\mu H(y) - H(y|x)) 
\end{eqnarray}

\section{Results}

\begin{figure*}[t]
\includegraphics[width=17cm, bb = 0 0 1450 620]{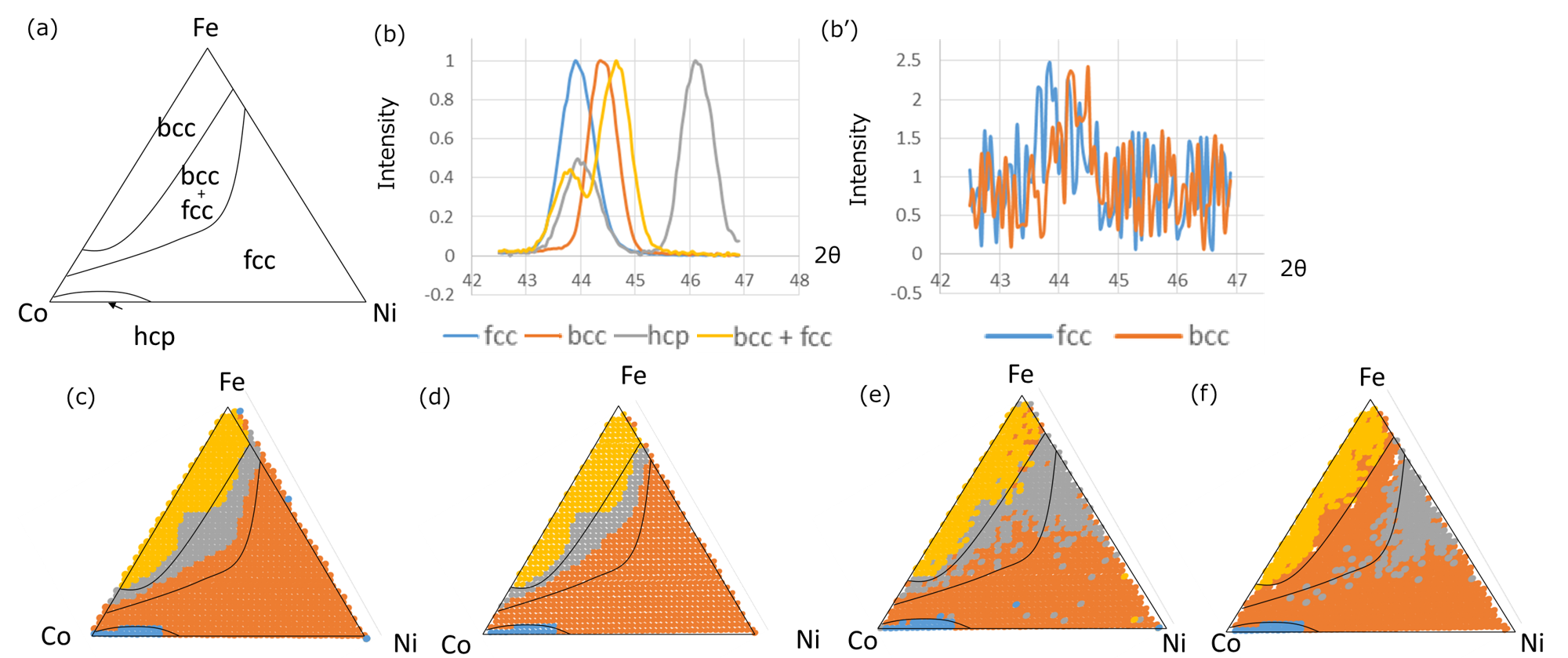}
\caption{Result of clustering of X-ray diffraction data of Fe-Co-Ni ternary-alloy thin film. (a) Phase map deduced from individual XRD patterns of spread wafer. (b) Example of XRD patterns where random noise was added to the diffraction data. (b') Example of XRD patterns where random noise was added. (c,d)  Result of cluster analysis using (c) IMSAT ($V_{y} = 4$) and (d) NC-DTW. (e,f) Result of cluster analysis using (e) IMSAT ($V_{y} = 4$) and (f) NC-DTW where random noise was added to the diffraction data.}
\label{fig:trialloy}
\end{figure*}

We applied our algorithm to the clustering of a line chart. Figure.\ \ref{fig:trialloy} (a) shows the phase map deduced from individual XRD patterns of Fe-Co-Ni ternary-alloy thin film by hand \cite{iwasaki2017}. The number of data $N$ is 1240. There are four types of the diffraction data, fcc, bcc, hcp and mixture of fcc and bcc. Examples of XRD patterns are shown in Fig.\ref{fig:trialloy} (b). We used commonly reported parameter values for neural networks. We set the network dimensionality to $d$-1200-1200-$V_{y}$ for clustering, where $d$( = 89) is input dimensionality. $N_{l}$, $\mu$ and $\lambda$ were set as 0 (unsupervised learning), $0.2$ and $0.2$, respectively. We set the size of the mini-batch to 64 and ran 50 epochs. We also tried the clustering using NC-DTW. We used the same parameters as Iwasaki's paper for NC-DTW. We set the window size $w$ to be 10 (0.5 degrees) and used hierarchy clustering analysis with average linkage method. The automated composition-phase maps using IMSAT and NC-DTW are shown in Fig. \ref{fig:trialloy} (c,d). A phase map using IMSAT and NC-DTW appears nealy the same.

We also examined the robustness against the noise of IMSAT and NC-DTW. Figure.\ \ref{fig:trialloy} (b') shows examples of XRD patterns where random noise was added to the diffraction data. One can see that the XRD patterns are noisy and difficult to classify by hand. Figure.\ \ref{fig:trialloy} (e,f) shows automated composition-phase maps using (e) IMSAT and (f) NC-DTW. Surprisingly, IMSAT succeeded in clustering of noisy XRD patterns and was more accurate than NC-DTW. It may originates from the fact that IMSAT can cancel out the noise inside neural network.



\begin{figure}[t]
\includegraphics[width=8cm, bb = 0 0 645 680]{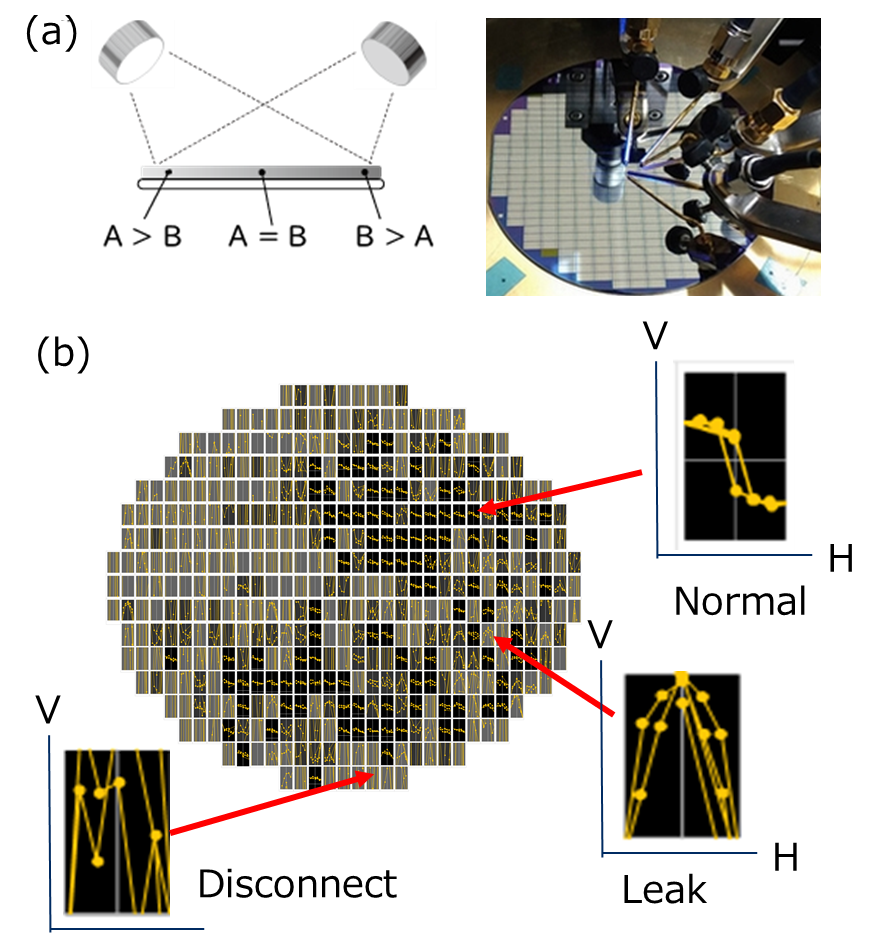}
\caption{Magnetic thin film fabricated by composition spread sputtering (a) and the hysteresis curve of ANE where thermo electric voltage exhibits a hysteresis curve depending on the external magnetic field (b). Figure.\ \ref{fig:fe} (b) also shows examples of the thermoelectric voltage curve of the disconnected sample and the leaked sample.  We measured the thermoelectric voltages of the thin film using semi-automatic wafer prober \cite{apollowwave}.}
\label{fig:fe}
\end{figure}

To verify the versatility, we also applied IMSAT to the clustering of scatter graph data; clustering of the hysteresis curve of magnetic FePt thin film. The FePt thin film was fabricated by composition spread sputtering. Figure.\ \ref{fig:fe} shows the example of the thin film fabricated by composition spread sputtering (a) and the hysteresis curve of anomalous Nernst effect (ANE) where thermo electric voltage exhibits a hysteresis curve depending on the external magnetic field (b) \cite{miyasato2007, gerrit2012}. The shape of the curve changes if fabrication of the thin film failed. There are two reasons for the failure, disconnection inside the sample and the insulator basis leaking onto the sample. Figure.\ \ref{fig:fe} (b) also shows examples of the thermoelectric voltage curve of the disconnected sample and the leaked sample. Typical curves of the disconnected sample and the leaked sample are random noise and V-shaped curve.



To implement IMSAT, we used a parameters that were almost the same as the clustering of XRD patterns. We set the network dimensionality to $d$-1200-1200-$V_{y}$ for clustering, where $d$( = $28 \times 28$) is input dimensionality. $N_{l}$, $\mu$, $\lambda$ were set to be 0 (unsupervised learning), $0.2$ and $0.2$, respectively. We set the size of the mini-batch to 40 and ran 50 epochs.

The left column of Table. \ref{tab:ane} shows the results of the automatic clustering of the voltage curve of the ANE of FePt thin film using IMSAT. Accuracy was calculated by $N(r_{imsat}; r_{hand})/N(r_{hand})$ where $N(r_{hand})$ is the number of the samples that were classified as $r_{hand}$(normal, disconnect, leak) by the clustering by hand. Clustering by hand was implemented by considering the shape of curvature and the results of the four-terminal measurement. Clearly, our algorithm was successful and highly accurate in classifying the normal samples. However, accuracy of  the classification of the leaked samples and disconnected sample was not so high, possibly because leakage and disconnection can occur simultaneously.

\begin{table*}[htb]
\begin{tabular}{|c|r|r|r|r|r|r|}
\hline
 & \multicolumn{3}{c}{Results with unsupervised IMSAT} & \multicolumn{3}{|c|}{Results with semi-supervised IMSAT} \\
\hline
 & Normal & Disconnect & Leak & Normal & Disconnect & Leak \\
\hline
 Normal (by hand) & 95 (94.4\%) & 0 (0\%) & 2 (5.6\%)  & 95 (94.4\%) & 1 (2.8\%) & 1(2.8\%)    \\
\hline
 Disconnect (by hand) & 18 (6.4 \%)  & 165(59.5\%) & 95(34.1\%) & 4 (1.4\%)  & 239(82.3\%) & 35(12.5\%)  \\
\hline
 Leak (by hand) & 13(24.5\%) & 12 (22.6\%)  &28(52.8\%)  & 4(7.5\%) & 18(33.4\%)  & 31(57.0\%)  \\
\hline
$R_{pert}$ & \multicolumn{3}{c}{0.275} & \multicolumn{3}{|c|}{0.465} \\
\hline
\end{tabular}
\caption{Result of automatic clustering of the voltage curve of anomalous Nernst effect  of FePt thin film using IMSAT and semi-supervised IMSAT. Accuracy was calculated by $N(r_{imsat}; r_{hand})/N(r_{hand})$ where $N(r_{hand})$ was the number of samples classified as $r_{hand}$(normal, disconnect, leak) by clustering by hand.}
\label{tab:ane}
\end{table*}

In terms of industrialization, classifying a failure sample as normal sample is critical. The left column of Table. \ref{tab:ane} shows that IMSAT sometimes classifies a failure sample as a normal sample because IMSAT just tries to make data representation locally invariant. We addressed the problem with semi-supervised learning where a penalty is added to misclassification of labeled data. The samples for labeled data are randomly chosen from the samples which are classified as normal sample by IMSAT though they were classified as failure sample by hand. We set $N_{l}$ as $5$ and $\beta$ as $3.34$. The right column of Table. \ref{tab:ane} shows the result of automatic clustering using semi-supervised learning. Semi-supervised learning suppressed the mis-classification by adding a penalty though it decreased the local invariant ($R_{pert}$) at the same time. This indicates semi-supervised IMSAT can flexibly respond to user's needs by regarding small, specific differences as important. We could not achieve 100\% accuracy with normal sample, possibly because the amounts of leakage and disconnection was not discrete quantity.


\section{Conclusion}
We presented how IMSAT can effectively classify raw experimental data without hands-on searches of kernel functions or preparation of large amounts of data for supervised learning. We demonstrated the clustering of XRD patterns using unsupervised IMSAT and the thermoelectric curve using a semi-supervised IMSAT and we showed that IMSAT is versatile and robust against noise and easily tunable by small data supervising. Our algorithm will accelerate the automation of big data collection and open a way to study artificial intelligent driven material development.

This work was financially supported by JST-ERATO Grant Number JPMJER1402 and JST-PRESTO, grant number JPMJPR17N4.

\bibliography{swdbib}

\end{document}